\documentclass[11pt,twoside]{article}
\usepackage{asp2006}
\usepackage{epsf}
\usepackage{psfig}
\usepackage{lscape}

\begin{document}
\title{The Future of Technical Libraries}
\author{Michael J. Kurtz; Guenther Eichhorn, Alberto Accomazzi, Carolyn 
Grant, Edwin Henneken, Donna Thompson, Elizabeth Bohlen, and Stephen S. Murray}
\affil{Harvard-Smithsonian Center for Astrophysics}

\begin{abstract}

Technical libraries are currently experiencing very rapid change.  In
the near future their mission will change, their physical nature will
change, and the skills of their employees will change.  While some
will not be able to make these changes, and will fail, others will
lead us into a new era.

\end{abstract}

\section{Introduction}

We are at the end of the paper era.  For the past two centuries
libraries have been primarily devoted to building and maintaining vast
collections of paper, documenting the myriad intellectual activities
of mankind.  These activities are nearing the end of their useful
lives.

Many existing libraries will not be able to meet the challenges, and
they will close.  Others will provide the intellectual leadership
necessary to support the new information seeking and storing paradigm,
and will thrive.

\section{The Problem}

Technical libraries, unlike more general purpose libraries, are not
supposed to be fun.  Typical users are professionals who are engaged
in a work function.  In astronomy the major use has been to read
journal articles, with catalogs, atlases, conference proceedings, and
monographs secondary uses.

Today nearly all these sources are original electronic; and nearly all
of the historical journal material, and much of the older rest has
already been transformed into electronic form.  The convenience of
finding and using these materials at one's desk is so great that the
physical use of the physical technical library has virtually
collapsed.  Many small departmental libraries have already closed.

While many libraries maintain responsibility for what to buy, this is
also rapidly ending.  Economic trends in the publishing industry are
driving libraries to band together to form large buying consortia to
negotiate with large publishers (or consortia of small publishers)
for comprehensive packages of journals and other publications.  This
effectively removes the responsibility for collection development from
individual libraries.  The adoption of open access would also have
this effect.

Finally new search technologies, such as provided by SIMBAD, NED, ADS,
DataScope, Google, Amazon, and others now provide a very sophisticated
reference desk function at the desktop.

It is clear that major changes will be required for libraries to meet
these challenges.

\section{The Future}

The future is now!

Astronomy is a leader in developing the new technical libraries.
Astronomers have built several new, large libraries in the last 20
years, essentially defining the new paradigm.

Obviously ADS, CDS, and NED are all basically libraries, but HEASARC,
MAST, the ESO archive, the Chandra archive, etc are also,
fundamentally, digital libraries.  The same technological changes
which are making paper obsolete are also now blurring the distinction
between a library and an archive.  Library based information systems,
such as D-Space and Fedora make this trend even more clear.

The new library structure in astronomy is based on the existence of
deep interlinking between the various groups.  This is enabled by rich
sets of meta-data.

An example of this interlinking is the connection between ADS and the
MAST based on a research article indexed in ADS having used data from
the HST and stored in the MAST.  The descriptive meta-data which
allows this link was first developed by Sarah Stevens-Rayburn.  Many
archives now create meta-data of this sort; ADS provides thousands of
these links.

In the future the creation and curation of meta-data will be of even
greater importance.  Already international efforts, via the virtual
observatory efforts in many countries, and the International Virtual
Observatory Alliance, are creating new, complex meta-data standards
for astronomical data of various types.  There will need to be
organizations to implement and maintain the large meta-data sets which
will come out of the VO work.  Libraries are those organizations.

\section{Conclusions}

The word library comes from a word for a kind of early paper.  As we
move inexorably away from paper as the medium for the transmission of
ideas the nature of libraries {\it per force} will change.  New
talents will be required to manage the new situations.  Librarians
will no longer need to bind paper volumes, but they will need to write
parsers.

The library of the future will be even more important to the practice
of research than is true currently.  Far from being quaint backwaters
of aging paper our libraries will be those organizations which index,
organize, and make sense of our massive, highly heterogeneous digital
data flows.

\end{document}